\documentclass[english,11pt,nofootinbib,showpacs,notitlepage,groupedaddress,superscriptaddress]{revtex4-1}
\usepackage[T1]{fontenc}
\usepackage{babel}
\usepackage[utf8]{inputenc}
\usepackage{amssymb,mathrsfs,amsfonts,amsmath,euscript,textcomp,graphicx,wrapfig,color,colortbl,txfonts,lmodern,multirow,bigdelim,dsfont,array,longtable}
\usepackage[section]{placeins}
\usepackage[usenames,dvipsnames,svgnames,table]{xcolor}
\definecolor{darkgreen}{rgb}{0,.7,0}
\usepackage[colorlinks,linkcolor=blue,citecolor=darkgreen,pagebackref=true]{hyperref}
\usepackage{float}
\def\[{\left[}
\def\]{\right]}
\def\({\left(}
\def\){\right)}

\def\1{{\bf CI}}
\def\2{{\bf CII}}
\def\3{{\bf CIII}}

\def\r{\mathbb{R}}

\newcommand{\eq}[1]{\begin{equation}#1\end{equation}}

\newcommand{\diag}{\mathop{\rm diag}\nolimits}

\newcommand{\lrp}[1]{\left(#1\right)}

\newcommand{\nd}{\noindent}

\newenvironment{tightcenter}{%
  \setlength\topsep{0pt}
  \setlength\parskip{0pt}
  \begin{center}
}{%
  \end{center}
}

\DeclareGraphicsExtensions{.pdf,.png,.jpg}

 \catcode`\@=11 \@addtoreset{equation}{section}\catcode`\@=12
\newcommand{\R}{ {\mathbb R} }

\begin{document}

\title{On stable  exponential cosmological  solutions   in the EGB  model with a $\Lambda$-term
 in dimensions $D = 5,6,7,8$}

\author{Chirkov D.M.}
\affiliation{Bauman Moscow State Technical University, 2-ya Baumanskaya ul., 5, Moscow, 105005, Russia}

\author{Toporensky A.V.}
\affiliation{Sternberg Astronomical Institute, Lomonosov Moscow State University,
Universitetsky Prospect, 13, Moscow 119991, Russia}
\affiliation{Kazan Federal University, Kremlevskaya 18, Kazan 420008, Russia}

\begin{abstract}

A $D$-dimensional  Einstein-Gauss-Bonnet (EGB) flat cosmological model with a cosmological term $\Lambda$ is considered. We focus on solutions with exponential dependence of scale factor on time. Using previously developed general analysis of stability of such solutions done by V.D.Ivashchuk (2016) we apply the criterion from that paper to all known exponential solutions upto dimensionality 7+1. We show that this criterion which guarantees stability of solution under consideration is fulfilled for all combination of coupling constant of the theory
except for some discrete set.

\end{abstract}

\maketitle


\section{Introduction}

Lovelock gravity \cite{Low} can be considered as the most conservative modification of General Relativity (GR) in the
sense that equation of motion of this theory are second order differential equations (the same as in GR)
in contrast to other metric theories, usually leading to fourth order equations (though some other
approaches with the same property exist, for example, Palatini version of $f(R)$ theory or $f(T)$ theory).
Usually increasing of the order of equations leads to a variety of new solutions, some of them without
a GR limit (for example, famous "false radiation" vacuum isotropic solution in $f(R)$), Lovelock gravity
being the second order theory is free from this solution. However, due to rather complicated equation
of motion Lovelock theory can also contain some solutions without a GR analog. One of such example
is exponential solutions in anisotropic cosmology.

In GR there is unique vacuum solution for a flat anisotropic Universe -- the Kasner solution (which,
strictly speaking is a one-dimensional set of solutions). Scale factors of this solution have a power-law
behavior in time. A version of power-law solution is known for general Lovelock gravity. However,
when higher Lovelock terms (starting from the Gauss-Bonnet term) are taken into account, new type
of solution with exponential time behavior of scale factors (i.e. constant Hubble parameters) appears.
Such solutions exist also for non-vacuum Universe, and belong to two different cases. If matter content
is different from the cosmological constant, the solution exists only in a very special case of the Universe with a constant
volume. For the matter in the form of a cosmological constant there is no restriction
on the volume. In this latter case all exponential solutions appear to be a subject of rather strict
condition: a space is divided into a restricted number (for Gauss-Bonnet theory, maximum three) of
isotropic subspaces. The fact that this division is not introduced "by hand" and appears naturally
from equations of motion makes exponential solution interesting for model building in multidimensional
cosmology. Any application should be preceded by studies of stability. Stability of exponential
solution have been considered recently in several papers, and, in particulary, it was shown that
in Einstein - Gauss-Bonnet theory the necessary condition for stability is volume increasing.
As for sufficient condition, a special algebraic relation should be satisfied.

In this paper we consider $D$-dimensional gravitational model
with Gauss-Bonnet term and cosmological term $\Lambda$.
The goal of the present paper is to check explicitly this relation for all known exponential
solution up to seven space dimensions. We note that so-called Gauss-Bonnet term appeared in string theory as a correction to the (Fradkin-Tseytlin) effective action \cite{Zwiebach}-\cite{MTs}.

 We note that at present the Einstein-Gauss-Bonnet (EGB) gravitational model and  its modifications,
see   \cite{Ishihara}-\cite{Ivas-16-2} and refs. therein,  are intensively studied in  cosmology,
e.g. for possible explanation  of  accelerating  expansion of the Universe which follow from supernovae (type Ia) observational data \cite{Riess,Perl,Kowalski}. Another applications are related to  Gauss-Bonnet-AdS black holes and holographic
description of certain quantum systems (e.g. superfluids), see \cite{BLMSY,KonZh} and refs. therein.

Here we consider examples of solutions in dimensions $D = 1 + d$ = $5,6,7,8$.
We  study   the stability of these solutions in a class of cosmological solutions with diagonal metrics by using results of  refs. \cite{ErIvKob-16,Ivas-16}, see also approach of ref. \cite{Pavl-15}.

Several sets of special stable  exponential solutions with zero variation of the effective gravitational constant
for two and three factor spaces were found recently in refs.  \cite{Ivas-16-2,Ern-Ivas-16-3} and \cite{Ern-Ivas-16-4},
respectively. It should be noted that  two  special solutions from ref. \cite{Ern-Ivas-16-3} for $D = 22, 28$ and $\Lambda = 0$ were found earlier in ref.  \cite{IvKob}. In ref. \cite{ErIvKob-16} it was proved that these solutions are stable.

\section{The set up}

The action of the model reads
\begin{equation}
  S =  \int_{M} d^{D}z \sqrt{|g|} \{ \alpha_1 (R[g] - 2 \Lambda) +
              \alpha_2 {\cal L}_2[g] \},
 \label{1}
\end{equation}
where $g = g_{MN} dz^{M} \otimes dz^{N}$ is the metric defined on
the manifold $M$, ${\dim M} = D$, $|g| = |\det (g_{MN})|$, $\Lambda$ is
the cosmological term, $${\cal L}_2 = R_{MNPQ} R^{MNPQ} - 4 R_{MN} R^{MN} +R^2$$
is the standard Gauss-Bonnet term and  $\alpha_1$, $\alpha_2$ are
nonzero constants.

We consider the manifold
\begin{equation}
   M = \R  \times   M_1 \times \ldots \times M_n \label{2.1}
\end{equation}
with the metric
\begin{equation}
   g= - d t \otimes d t  +
      \sum_{i=1}^{n} B_i e^{2v^i t} dy^i \otimes dy^i,
  \label{2.2}
\end{equation}
where   $B_i > 0$ are arbitrary constants, $i = 1, \dots, n$, and
$M_1, \dots,  M_n$  are one-dimensional manifolds (either $\R$ or $S^1$)
and $n > 3$.

Equations of motion for the action (\ref{1})
give us the set of  polynomial equations \cite{ErIvKob-16}
\begin{eqnarray}
  G_{ij} v^i v^j + 2 \Lambda
  - \alpha   G_{ijkl} v^i v^j v^k v^l = 0,  \label{2.3} \\
    \left[ 2   G_{ij} v^j
    - \frac{4}{3} \alpha  G_{ijkl}  v^j v^k v^l \right] \sum_{i=1}^n v^i
    - \frac{2}{3}   G_{ij} v^i v^j  +  \frac{8}{3} \Lambda = 0,
   \label{2.4}
\end{eqnarray}
$i = 1,\ldots, n$, where  $\alpha = \alpha_2/\alpha_1$. Here
\begin{equation}
G_{ij} = \delta_{ij} -1, \qquad   G_{ijkl}  = G_{ij} G_{ik} G_{il} G_{jk} G_{jl} G_{kl}
\label{2.4G}
\end{equation}
are, respectively, the components of two  metrics on  $\R^{n}$ \cite{Iv-09,Iv-10}.
The first one is a 2-metric and the second one is a Finslerian 4-metric.
For $n > 3$ we get a set of forth-order polynomial  equations.

We note that for $\Lambda =0$ and $n > 3$ the set of equations (\ref{2.3}) and (\ref{2.4}) has an isotropic solution $v^1 = \ldots = v^n = H$ only if $\alpha  < 0$ \cite{Iv-09,Iv-10}.
This solution was generalized in \cite{ChPavTop} to the case $\Lambda \neq 0$.

It was shown in \cite{Iv-09,Iv-10} that there are no more than
three different  numbers among  $v^1,\dots ,v^n$ when $\Lambda =0$. This is valid also
for  $\Lambda \neq 0$ if $\sum_{i = 1}^{n} v^i \neq 0$  \cite{Ivas-16}.

\section{Conditions for stability}

Here, as in \cite{ErIvKob-16,Ivas-16}, we deal with exponential solutions (\ref{2.2})
with non-static volume factor, which is proportional to $\exp(\sum_{i = 1}^{n} v^i t)$, i.e. we put
\begin{equation}
  K = K(v) = \sum_{i = 1}^{n} v^i \neq 0.
  \label{4.1}
\end{equation}

We  put the following restriction
\begin{equation}
  ({\rm R }) \quad  \det (L_{ij}(v)) \neq 0.
  \label{4.2}
\end{equation}
on the matrix
\begin{equation}
L =(L_{ij}(v)) = (2 G_{ij} - 4 \alpha G_{ijks} v^k v^s).
   \label{4.1b}
 \end{equation}

For general cosmological setup with the metric
\begin{equation}
 g= - dt \otimes dt + \sum_{i=1}^{n} e^{2\beta^i(t)}  dy^i \otimes dy^i,
 \label{4.3}
\end{equation}
we have  the (mixed) set of algebraic and differential equations \cite{Iv-09,Iv-10}
\begin{eqnarray}
     E = G_{ij} h^i h^j + 2 \Lambda  - \alpha G_{ijkl} h^i h^j h^k h^l = 0,
         \label{4.3.1} \\
         Y_i =  \frac{d L_i}{dt}  +  (\sum_{j=1}^n h^j) L_i -
                 \frac{2}{3} (G_{sj} h^s h^j -  4 \Lambda) = 0,
                     \label{4.3.2a}
          \end{eqnarray}
where $h^i = \dot{\beta}^i$,
 \begin{equation}
  L_i = L_i(h) = 2  G_{ij} h^j
       - \frac{4}{3} \alpha  G_{ijkl}  h^j h^k h^l
       \label{4.3.3},
 \end{equation}
 $i = 1,\ldots, n$.

It was proved in \cite{Ivas-16} that a fixed point solution
$(h^i(t)) = (v^i)$ ($i = 1, \dots, n$; $n >3$) to eqs. (\ref{4.3.1}), (\ref{4.3.2a})
obeying restrictions (\ref{4.1}), (\ref{4.2}) is  stable under perturbations
\begin{equation}
 h^i(t) = v^i +  \delta h^i(t),
\label{4.3h}
\end{equation}
 $i = 1,\ldots, n$,  (as $t \to + \infty$)  if
\begin{equation}
  K(v) = \sum_{k = 1}^{n} v^k > 0
 \label{4.1a}
\end{equation}
and  it is unstable (as $t \to + \infty$) if $K(v) = \sum_{k = 1}^{n} v^k < 0$.

We remind the reader that the  perturbations $\delta h^i$
obey (in linear approximation) the following set of  equations
\cite{ErIvKob-16,Ivas-16}
 \begin{eqnarray}
   C_i(v) \delta h^i = 0, \label{4.2C} \\
   L_{ij}(v) \delta \dot{h}^j =  B_{ij}(v) \delta h^j,
  \label{4.3LB}
 \end{eqnarray}
  where
 \begin{eqnarray}
 C_i(v)  =  2 v_{i} - 4 \alpha G_{ijks}  v^j v^k v^s, \label{4.3.4} \\
 B_{ij}(v) = - (\sum_{k = 1}^n v^k)  L_{ij}(v) - L_i(v) + \frac{4}{3} v_{j},
                     \label{4.3.6}
 \end{eqnarray}
 $v_i = G_{ij} v^j$,
 and $i,j,k,s = 1, \dots, n$.

It was proved in ref. \cite{Ivas-16} that  the set of linear equations
on  perturbations (\ref{4.2C}), (\ref{4.3LB})  has the following solution
   \begin{eqnarray}
       \delta h^i = A^i \exp( - K(v) t ),
       \label{4.16}  \\
         \sum_{i =1}^{n} C_i(v)  A^i =0,
         \label{4.16A}
   \end{eqnarray}
    $i = 1, \dots, n$, when restrictions (\ref{4.1}), (\ref{4.2}) are imposed.

 It was shown also in~\cite{Ivas-16} that in the case when we have two different Hubble-like parameters $H, h$, i.e. when the vector $v=(\underbrace{H,H,H,H\ldots,H}_{m},\overbrace{h,\ldots,h}^{l})$,
and $K = m H + l h \neq 0$, the  matrix $L$ has a block-diagonal form: $L=\diag(L_{\mu\nu},L_{\alpha\beta}),$ where
\eq{L_{\mu\nu}=(1-\delta_{\mu\nu})(2+4\alpha S_{HH}),\;\; L_{\alpha\beta}=(1-\delta_{\alpha\beta})(2+4\alpha S_{hh}),\quad \mu,\nu=1\ldots m,\;\;\alpha,\beta=m+1,\ldots,n}
and $S_{HH},S_{hh}$ are functions of $H,h,m,l$. From this we immediately get that
if $h$ corresponds to 1-dimensional subspace, i.e. $l=1$, then $L_{\alpha\beta}$ is the $1\times 1$-block which equals to zero since $\delta_{\alpha\beta}=\delta_{m+1\,m+1}=1$ and $\det(L)=0$ in this case.

 Analogously, it was shown in~\cite{Ern-Ivas-16-4} that in the case when we have three different Hubble-like parameters $H, \mathcal{H}, h$, i.e. when the vector $v=(\underbrace{H,H,H,H\ldots,H}_{m},\overbrace{\mathcal{H},\ldots,\mathcal{H}}^{k_1},\overbrace{h,\ldots,h}^{k_2})$,
and $K = m H + k_1 {\cal H} +  k_2 h \neq 0$ the
 matrix $L$ has a block-diagonal form again: $L=\diag(L_{\mu\nu},L_{ab},L_{\alpha\beta})$ with
\eq{L_{\mu\nu}=(1-\delta_{\mu\nu})(2+4\alpha S_{HH}),\;\; L_{ab}=(1-\delta_{ab})(2+4\alpha S_{\mathcal{H}\mathcal{H}}),\;\; L_{\alpha\beta}=(1-\delta_{\alpha\beta})(2+4\alpha S_{hh})}
where $\mu,\nu=1\ldots m,\;\;a,b=m+1,\ldots,m+k_1,\;\;\alpha,\beta=m+k_1+1,\ldots,n$ and $S_{HH},S_{\mathcal{H}\mathcal{H}},S_{hh}$ are functions of $H,\mathcal{H},h,m,k_1,k_2$. If $h$ corresponds to 1-dimensional subspace then $\det(L)=0$ exactly for the same reason as in the previous case.

We will see particular examples of this situation in the next section. Moreover, the solutions in question
will leave this particular Hubble parameter $h$ unconstrained. From the continuity of $\det(L)$ as a function of $h$ we can conclude that in this case $\det(L)=0$ for all $h$, i.e. also when  $h$ is either coinciding with one of the other Hubble-like parameters ($H$ or  $H, \mathcal{H}$) or when the sum of  all  Hubble-like parameters $K$ is zero.

\section{Stability of fixed point solutions in \texorpdfstring{\MakeLowercase{d}}{}=4,5,6,7}
Now we apply the criterion of stability~(\ref{4.1a}) to (4+1)-,\,(5+1)-,\,(6+1)- and (7+1)-dimensional exponential solutions with non-static volume factor that have been obtained in~\cite{ChPavTop,ChPavTop1} and gather data concerning stability of these solutions in the tables~\ref{d=4}-\ref{d=7} below.

\subsection{d=4 and d=5}
Since the criterion works with restriction~(\ref{4.1}) only, first of all we evaluate determinant of the matrix~(\ref{4.1b}) for each solution and check if it equals to zero. In the case of singular matrix $L$ we can not say anything about stability of the corresponding solutions; such solutions are marked in tables~\ref{d=4} and~\ref{d=5} in gray.
\begin{table}[!h]
\begin{center}
\caption{Stability of solutions (d=4)}
\label{d=4}
  \begin{tabular}{|c|c|c|c|}
    \hline
    $(v^1,v^2,v^3,v^4)$ & Solution & $\det(L)$ & Stability condition \\
    \hline
    \multirow{2}{*}{$(H,H,h,h)$} &
    $\begin{array}{c}
     H^2=\frac{1}{4}\left(2\Lambda+\frac{1}{2\alpha}\pm\sqrt{\left(2\Lambda+\frac{1}{2\alpha}\right)^2-\frac{1}{\alpha^2}}\right),\;h=-\frac{1}{4\alpha H} \\
     \left\{\begin{array}{l}
       \Lambda>\frac{1}{4\alpha} \\
       \alpha>0
     \end{array}\right.\; \mbox{or}\;
     \left\{\begin{array}{l}
       \Lambda>-\frac{3}{4\alpha} \\
       \alpha<0
     \end{array}\right.
    \end{array}$ & ${\frac {256\,\left( {H}^{2}\alpha+1/4 \right) ^{4}}{{H}^{4}{\alpha}^{2}}}$ &
    $\begin{array}{c}
    \left\{\begin{array}{l}
       H\in(-\frac{1}{\sqrt{4\alpha}};0)\cup \\
       \cup(\frac{1}{\sqrt{4\alpha}};+\infty) \\
       \alpha>0
     \end{array}\right. \\ \mbox{or} \\
     H>0\;\;\mbox{if}\;\;\alpha<0
     \end{array}$ \\
    \cline{2-4}
    \rowcolor{gray!40}
    \cellcolor{white}& $H^2=h^2=-\frac{1}{4\alpha},\;\Lambda=-\frac{3}{4\alpha},\;\alpha<0$ & 0 & --- \\
    \hline
    \rowcolor{gray!40}
    \cellcolor{white} $(H,H,H,h)$ & $H^2=-\frac{1}{4\alpha},\;h\in\r,\;\Lambda=-\frac{3}{4\alpha},\;\alpha<0$ & 0 & --- \\
    \hline
    \multirow{2}{*}{$(H,H,H,H)$} & $\begin{array}{c}
                                     H^2=\frac{1}{4\alpha}\left(-1+\sqrt{1+\frac{4\alpha\Lambda}{3}}\right),\alpha>0,\Lambda>0 \\
                                     H^2=\frac{1}{4\alpha}\left(-1-\sqrt{1+\frac{4\alpha\Lambda}{3}}\right),\alpha<0,\Lambda<-\frac{3}{4\alpha} \\
                                     H^2=\frac{1}{4\alpha}\left(-1+\sqrt{1+\frac{4\alpha\Lambda}{3}}\right),\alpha<0,0<\Lambda<-\frac{3}{4\alpha} \\
                                     \mbox{\textbf{Vacuum solution:}}\;\; H^2=-\frac{1}{2\alpha},\;\alpha<0
                                   \end{array}$
    & $-12288\left( {H}^{2}\alpha+1/4 \right)^{4}$ & $H>0$ \\
    \cline{2-4}
    \rowcolor{gray!40}
    \cellcolor{white} & $H^2=-\frac{1}{4\alpha},\;\Lambda=-\frac{3}{4\alpha},\;\alpha<0$ & 0 & --- \\
    \hline
  \end{tabular}
\end{center}
\end{table}
\FloatBarrier\nd One can see that $\det(L)=0$ for all those solutions which exist for the single value of $\Lambda$ (given fixed $\alpha$); in all these cases $\alpha<0$. This is the solution with 3D isotropic subspace and one extra dimension (we already know from the previous section that the determinant is equal
to zero for this solution) and two particular cases of isotropic and 2D+2D solution. Note that the former
solution is in fact one-dimensional set of solution (because $h$ is a free parameter there), and two special
cases of other solutions with zero determinant appear to be particular points in this set. In this sense only
3D+1D solution (existing only for particular combination of coupling constants) has vanishing determinant.

\begin{table}[!h]
\begin{center}
\caption{Stability of solutions (d=5)}
\label{d=5}
  \begin{tabular}{|c|c|c|c|}
    \hline
    $(v^1,v^2,v^3,v^4,v^5)$ & Solution & $\det(L)$ & Stability condition \\
    \hline
    \rowcolor{gray!40}
    \cellcolor{white}$(H,H,-H,-H,h)$ &
    $H^2=\frac{1}{4\alpha},\;h\in\r,\;\Lambda=\frac{1}{4\alpha},\;\alpha>0$ & 0 & --- \\
    \hline
    \rowcolor{gray!40}
    \cellcolor{white} $(H,H,H,H,h)$ & $H^2=-\frac{1}{12\alpha},\;h\in\r,\;\Lambda=-\frac{5}{12\alpha},\;\alpha<0$ & 0 & --- \\
    \hline
    $(H,H,H,H,H)$ & $\begin{array}{c}
                                     H^2=\frac{1}{12\alpha}\left(-1+\sqrt{1+\frac{12\alpha\Lambda}{5}}\right),\alpha>0,\Lambda>0 \\
                                     H^2=\frac{1}{12\alpha}\left(-1-\sqrt{1+\frac{12\alpha\Lambda}{5}}\right),\alpha<0,
                                     \Lambda<-\frac{5}{12\alpha} \\
                                     H^2=\frac{1}{12\alpha}\left(-1+\sqrt{1+\frac{12\alpha\Lambda}{5}}\right),\alpha<0,0<\Lambda<-\frac{5}{12\alpha} \\
                                     \mbox{\textbf{Vacuum solution:}}\;\; H^2=-\frac{1}{6\alpha},\;\alpha<0
                                   \end{array}$
    & $-128\, \left( 12\,H^{2}\alpha+1 \right) ^{5}$ & $H>0$ \\
    \cline{2-4}
    \rowcolor{gray!40}
    \cellcolor{white} & $H^2=-\frac{1}{12\alpha},\;\alpha<0,\;\Lambda=-\frac{5}{12\alpha}$ & 0 & --- \\
    \hline
  \end{tabular}
\end{center}
\end{table}

The case $(H,H,H,h,h)$ with 3-dimensional isotropic subspace is more complicated, we consider it separately.

\begin{table}[!h]
  \centering
  \nd
  \begin{tabular}{p{9cm}@{\hskip .4cm}|@{\hskip .4cm}p{9cm}}
    \centering $\mathbf{\Lambda}$\textbf{-term solution} & \centering\arraybackslash \textbf{Vacuum solution} \\
    & \\
    $(\ast)\quad 192\,{H}^{6}{\alpha}^{3}-112\,{H}^{4}{\alpha}^{2}+\left( 64\Lambda{\alpha}+4\right){H}^{2}\alpha-1=0$
    $(\ast\ast)\quad h=-{\frac {4\,{H}^{2}\alpha+1}{8H\alpha}}$ \newline\newline
    It is easy to check that when $\alpha>0$ Eqs.~($\ast$)-($\ast\ast$) has at least one solution
     for any $\Lambda$;
    when $\alpha<0$ Eqs.~($\ast$)-($\ast\ast$) has at least one solution iff $\Lambda\geqslant-\frac{5}{12\alpha}$.
    In this case \newline
    $\makebox[\linewidth]{$\det(L)=-\,{\frac {248832 \left( {H}^{2}\alpha-1/4 \right) ^{3} \left( {H}^{2}\alpha+1/12 \right) ^{5}}{{H}^{6}{\alpha}^{3}}}$}$\newline
    One can easily check that $\det(L)=0$ iff $H^2=\frac{1}{4\alpha}$ or $H^2=-\frac{1}{12\alpha}$;
    $H^2=\frac{1}{4\alpha}$ being the solution of Eqs.~($\ast$)-($\ast\ast$) is a particular case of the family $(H,H,-H,-H,h)$;
    $H^2=-\frac{1}{12\alpha}$ being the solution of Eqs.~($\ast$)-($\ast\ast$) is a particular case of the family $(H,H,H,H,h)$.
    A solution of Eqs.~($\ast$)-($\ast\ast$) is stable if
    $3H+2h>0\iff \frac{16H^2\alpha-1}{4H\alpha}>0\iff$
    $\iff\left\{\begin{array}{l}
    H>0,\;\;\alpha<0 \\
    H>\frac{1}{\sqrt{16\alpha}}\;\mbox{or}\;-\frac{1}{\sqrt{16\alpha}}<H<0,\;\;\alpha>0
    \end{array}\right.$ & $H_1=H_2=H_3=H,\;H_4=H_5=\xi H$ \newline $H^2=-\frac{\xi^2+6\xi+3}{12\alpha\xi(3\xi+2)}\Biggl|_{\xi=\frac{\sqrt[3]{10}}{3}-\frac{\sqrt[3]{100}}{6}-\frac{2}{3}\approx -0.722}, \;\alpha<0$ \newline\newline
    In this case \newline
    $\begin{array}{c}
    (\star)\quad\det(L)=-1769472\cdot \\
    \cdot\left(H^4\alpha^2\left(\xi^2+\frac{2}{3}\xi+\frac{1}{3}\right)-\frac{H^2\alpha}{36}(\xi^2-10\xi-3)+\frac{1}{72}\right)\cdot \\
    \cdot\left(\xi(\xi+2)H^2\alpha+\frac{1}{4}\right)\cdot\left(H^2\alpha+\frac{1}{12}\right)
    \end{array}$\newline It is easy to check that solution~($\star$) does not turn determinant $L$ to zero, so this solution is stable if $H>0$.
  \end{tabular}
\end{table}
\FloatBarrier\nd It is interesting to note that both $\Lambda$-term and vacuum solutions with 3D isotropic subspace are stable and stability condition requires expanding of this 3D subspace. As well as in the case $d=4$ we see that $\det(L)=0$ for all those solutions which exist for the single value of $\Lambda$ (given fixed $\alpha$). There are two one-dimensional set of solutions and a particular case of isotropic solution
when it coincides with a point in 4D+1D set.
Note also that all vacuum solutions both in $d=4$ and $d=5$ cases exist only for $\alpha<0$ and stable for $H>0$; there are no vacuum solutions with singular matrix $L$.

\subsection{d=6}
In this subsection we generalize the above results to the case $d=6$. There are two special solutions which exist for the single value of $\Lambda$ (given fixed $\alpha$) such that $\det(L)=0$:
\eq{(H,H,H,h,h,\mathds{h}):\quad H^2=\frac{1}{12\alpha},\;\;h=-2H,\;\;\mathds{h}\in\r,\;\;\Lambda=\frac{1}{4\alpha},\;\;\alpha>0\label{3+2+1}}
\eq{(H,H,H,H,H,h):\quad H^2=-\frac{1}{24\alpha},\;\;h\in\r,\;\;\Lambda=-\frac{5}{16\alpha},\;\;\alpha<0\label{5+1}}
Here we encounter the situation that is qualitatively different from what we see in (4+1)- and (5+1)-dimensional models: matrix $L$ is singular not only for one.dimensional sets of
solutions which exist for single value of $\Lambda$ and has free parameter $h$, but this matrix is turned to be singular for other solutions (however, still for some special values of $\Lambda$); in the table~\ref{d=6} for each family we write $(H,h,\Lambda)$ for solutions with two different Hubble parameters and $(H,\Lambda)$ for isotropic solution such that $\det(L)=0$. We do not write down these solutions itself due to their awkwardness and describe each family by its splitting onto isotropic subspaces; the reader can find formulas in~\cite{ChPavTop,ChPavTop1}. Some solutions listed in the table~\ref{d=6} are particular cases of solutions~(\ref{3+2+1})-(\ref{5+1}); we point out overlapping families in the last column of the table~\ref{d=6}. There are two solutions (both have 3-dimensional isotropic subspaces) that do not overlap with solutions~(\ref{3+2+1}) and~(\ref{5+1}) but have nevertheless vanishing determinant; we highlight these special solutions in the table~\ref{d=6} by gray color.
\begin{table}[!h]
\caption{d=6: special points of solutions}\label{d=6}
  \begin{tabular}{|c|c|c|}
    \hline\centering
    Family of solutions & with $(H,h,\Lambda)$ such that $\det(L)=0$ & overlap to \\
    \hline
    \multirow{5}{*}{$(H,H,H,h,h,h)$}
      & \cellcolor{gray!40} $\lrp{\pm\frac{1}{\sqrt{-24\alpha}},\mp\frac{5}{\sqrt{-24\alpha}},-\frac{77}{16\alpha}}$ & \cellcolor{gray!40} nothing \\
      \cline{2-3}
      & \cellcolor{gray!40} $\lrp{\pm\frac{1}{\sqrt{3\alpha}},\mp\frac{7}{\sqrt{12\alpha}},-\frac{143}{4\alpha}}$ & \cellcolor{gray!40} nothing \\
      \cline{2-3}
      & $\lrp{\pm\frac{1}{\sqrt{3\alpha}},\mp\frac{1}{\sqrt{12\alpha}},\frac{1}{4\alpha}}$ & $(H,H,H,h,h,\mathds{h})$ \\
      \cline{2-3}
      & $\lrp{\pm\frac{1}{\sqrt{12\alpha}},\mp\frac{1}{\sqrt{3\alpha}},\frac{1}{4\alpha}}$ & $(H,H,H,h,h,\mathds{h})$ \\
      \cline{2-3}
      & $\lrp{\pm\frac{1}{\sqrt{-24\alpha}},\pm\frac{1}{\sqrt{-24\alpha}},-\frac{5}{16\alpha}}$ & $(H,H,H,H,H,h)$ \\
    \hline
    \multirow{2}{*}{$(H,H,H,H,h,h)$} & $\lrp{\pm\frac{1}{\sqrt{12\alpha}},\mp\frac{1}{\sqrt{3\alpha}},\frac{1}{4\alpha}}$ & $(H,H,H,h,h,\mathds{h})$ \\
    \cline{2-3}
    & $\lrp{\pm\frac{1}{\sqrt{-24\alpha}},\pm\frac{1}{\sqrt{-24\alpha}},-\frac{5}{16\alpha}}$ & $(H,H,H,H,H,h)$ \\
    \hline
    $(H,H,H,H,H,H)$ & $\lrp{\pm\frac{1}{\sqrt{-24\alpha}},-\frac{5}{16\alpha}}$ & $(H,H,H,H,H,h)$ \\
    \hline
  \end{tabular}
\end{table}

\FloatBarrier\nd Note that as well as in the cases of $d=4$ and $d=5$ there are no vacuum solutions with $\det(L)=0$.

\subsection{d=7}
In this subsection we generalize the above results to the case $d=7$. There are three special solutions which exist for the single value of $\Lambda$ (given fixed $\alpha$) such that $\det(L)=0$:
\eq{(H,H,H,H,h,h,\mathds{h}):\quad H^2=\frac{1}{24\alpha},\;\;h=-3H,\;\;\mathds{h}\in\r,\;\;\Lambda=\frac{13}{48\alpha},\;\;\alpha>0\label{4+2+1}}
\eq{(H,H,H,-H,-H,-H,h):\quad H^2=\frac{1}{8\alpha},\;\;h\in\r,\;\;\Lambda=\frac{3}{16\alpha},\;\;\alpha>0\label{3-3+1}}
\eq{(H,H,H,H,H,H,h):\quad  H^2=-\frac{1}{40\alpha},\;\;h\in\r,\;\;\Lambda=-\frac{21}{80\alpha},\;\;\alpha<0\label{6+1}}
The other families of solutions have only special solutions with $\det(L)=0$; in the table~\ref{d=7} for each family we write $(H,h,\mathds{h},\Lambda)$ for solutions with three different Hubble parameters, $(H,h,\Lambda)$ for solutions with two different Hubble parameters and $(H,\Lambda)$ for isotropic solution. As well as in the previous case we do not write down solutions itself due to their awkwardness, all formulas can be found in~\cite{ChPavTop,ChPavTop1}. As in the $d=6$ case we list overlapping families in the last column of the table~\ref{d=7}. We see again that there are several solutions with 3D isotropic subspaces that do not overlap solutions~(\ref{4+2+1})-(\ref{6+1}); we highlight these solutions by gray color.
\begin{table}[!h]
\caption{d=7: special points of solutions}\label{d=7}
  \begin{tabular}{|c|c|c|}
    \hline\centering
    Family of solutions & with $(H,h,\mathds{h},\Lambda)$ such that $\det(L)=0$ & overlap to \\
    \hline
    \multirow{2}{*}{$(H,H,H,h,h,\mathds{h},\mathds{h})$} & $\lrp{\pm\frac{1}{\sqrt{8\alpha}},\mp\frac{1}{\sqrt{8\alpha}},\mp\frac{1}{\sqrt{8\alpha}},\frac{3}{16\alpha}}$ & $(H,H,H,-H,-H,-H,h)$ \\
    \cline{2-3}
    & $\begin{array}{c}
         \lrp{\pm\frac{1}{\sqrt{24\alpha}},\mp\frac{\sqrt{3}}{\sqrt{8\alpha}},\pm\frac{1}{\sqrt{24\alpha}},\frac{13}{48\alpha}} \\
         \lrp{\pm\frac{1}{\sqrt{24\alpha}},\pm\frac{1}{\sqrt{24\alpha}},\mp\frac{\sqrt{3}}{\sqrt{8\alpha}},\frac{13}{48\alpha}}
       \end{array}$ & $(H,H,H,H,h,h,\mathds{h})$ \\
    \hline
    \multirow{5}{*}{$(H,H,H,H,h,h,h)$}
      & \cellcolor{gray!40} $\lrp{\pm\frac{1}{\sqrt{-40\alpha}},\pm\frac{7}{\sqrt{-40\alpha}},-\frac{2409}{400\alpha}}$ & \cellcolor{gray!40} nothing \\
      \cline{2-3}
      & \cellcolor{gray!40} $\lrp{\pm\frac{1}{\sqrt{8\alpha}},\mp\frac{5}{\sqrt{8\alpha}},-\frac{285}{16\alpha}}$ & \cellcolor{gray!40} nothing \\
      \cline{2-3}
      & $\lrp{\pm\frac{1}{\sqrt{24\alpha}},\mp\frac{\sqrt{3}}{\sqrt{8\alpha}},\frac{13}{48\alpha}}$ & $(H,H,H,H,h,h,\mathds{h})$ \\
      \cline{2-3}
      & $\lrp{\pm\frac{1}{\sqrt{-40\alpha}},\mp\frac{1}{\sqrt{-40\alpha}},-\frac{21}{80\alpha}}$ & $(H,H,H,H,H,H,h)$ \\
      \cline{2-3}
      & $\lrp{\pm\frac{1}{\sqrt{8\alpha}},\mp\frac{1}{\sqrt{8\alpha}},\frac{3}{16\alpha}}$ & $(H,H,H,-H,-H,-H,h)$ \\
    \hline
    \multirow{2}{*}{$(H,H,H,H,H,h,h)$}
    & $\lrp{\pm\frac{1}{\sqrt{24\alpha}},\mp\frac{\sqrt{3}}{\sqrt{8\alpha}},\frac{13}{48\alpha}}$ & $(H,H,H,H,h,h,\mathds{h})$ \\
    \cline{2-3}
    & $\lrp{\pm\frac{1}{\sqrt{-40\alpha}},\pm\frac{1}{\sqrt{-40\alpha}},-\frac{21}{80\alpha}}$ & $(H,H,H,H,H,H,h)$ \\
    \hline
    $(H,H,H,H,H,H,H)$ & $\lrp{\pm\frac{1}{\sqrt{-40\alpha}},-\frac{21}{80\alpha}}$ & $(H,H,H,H,H,H,h)$ \\
    \hline
  \end{tabular}
\end{table}
\FloatBarrier\nd There are no (7+1)-dimensional vacuum solutions with $\det(L)=0$ again.

\section{Conclusions}

We have considered the  $D$-dimensional  Einstein-Gauss-Bonnet (EGB) model
with the $\Lambda$-term and two constants $\alpha_1$ and $\alpha_2$.
The full list of the solutions with exponential time dependence of the scale factors have been found in
\cite{ChPavTop, ChPavTop1},
here we consider stability of these solutions. As it have been described in \cite{Ivas-16} the necessary condition for
stability is the condition that the overall volume of the space considered is growing. Checking  this
condition on known solution is straightforward. The sufficient condition for stability is more cumbersome
to check. In the present paper we have checked this condition for all known exponential solutions up
to dimension $7+1$.

For summarizing the results obtained it is worth to remember that exponential solutions can be divided
into two groups. Solutions of the first group (we can call them as special solutions) exist only for
coupling satisfying an additional linear relation. On the other hand, one of Hubble parameters of the
solution remained unconstrained. Our results shown that all such solution considered in the present paper
do not satisfy the sufficient condition for stability, and, so, the dynamics in the vicinity of these
solutions requires further investigation.

One the contrary, the second group of solutions (existing for non-zero measure of possible couplings
and with all Hubble parameters fully determined) satisfy the sufficient condition for the stability
except for very few special sets of couplings.
One of the reason for this situation may be the fact that branches of this second group of solution can intersect
with branches of the special solutions. On the other hand, we identified several particular coupling
which do not satisfy the sufficient condition for stability and do not originate from intersection with
any other branches of solutions. Curiously, such a situation occurs only for solutions with 3D isotropic
subspace.

To conclude in brief our results shows that (at list up to the dimension 7+1) the stability of exponential
solution with growing spatial volume in Gauss-Bonnet cosmology can not be proved in the linear perturbation analysis only for
a discrete set of couplings. What happens in this case (when the necessary condition for stability is fulfilled and the sufficient condition is not) needs further analysis.

\begin{acknowledgments}
The work of A.T. is supported by RFBR grant 17-02-01008 and
by the Russian Government
Program of Competitive Growth of Kazan Federal University. Authors are grateful to Vladimir Ivashchuk
for discussions.
\end{acknowledgments}

\end{document}